\documentclass[prb,superscriptaddress,twocolumn,showkeys,showpacs]{revtex4}
\usepackage{graphicx}
\begin{document}
\author{H. Schneider}
\email{schneiho@uni-mainz.de}
\affiliation{Institut f\"ur Physik, Johannes Gutenberg-Universit\"at, 55099 Mainz, Germany}
\author{G. Jakob}
\affiliation{Institut f\"ur Physik, Johannes Gutenberg-Universit\"at, 55099 Mainz, Germany}
\author{M. Kallmayer}
\affiliation{Institut f\"ur Physik, Johannes Gutenberg-Universit\"at, 55099 Mainz, Germany}
\author{H.~J. Elmers}
\affiliation{Institut f\"ur Physik, Johannes Gutenberg-Universit\"at, 55099 Mainz, Germany}
\author{M. Cinchetti}
\affiliation{Technische Universit\"at Kaiserslautern, 67663 Kaiserslautern, Germany}
\author{B. Balke}
\affiliation{Institut f\"ur Anorganische und Analytische Chemie, Johannes Gutenberg-Universit\"at, 55099 Mainz, Germany}
\author{S. Wurmehl}
\affiliation{Institut f\"ur Anorganische und Analytische Chemie, Johannes Gutenberg-Universit\"at, 55099 Mainz, Germany}
\author{C. Felser}
\affiliation{Institut f\"ur Anorganische und Analytische Chemie, Johannes Gutenberg-Universit\"at, 55099 Mainz, Germany}
\author{M. Aeschlimann}
\affiliation{Technische Universit\"at Kaiserslautern, 67663 Kaiserslautern, Germany}
\author{H. Adrian}
\affiliation{Institut f\"ur Physik, Johannes Gutenberg-Universit\"at, 55099 Mainz, Germany}
\title{Epitaxial film growth and magnetic properties of $\mathbf{Co_2FeSi}$}
\date{\today}
\begin{abstract}
We have grown thin films of the Heusler compound $\rm Co_2FeSi$ by RF magnetron sputtering.  On (100)-oriented MgO substrates we find fully epitaxial (100)-oriented and $\rm L2_1$ ordered growth. On $\rm{Al_2O_3\,(11\bar{2}0)}$ substrates, the film growth is (110)-oriented, and several in-plane epitaxial domains are observed. The temperature dependence of the electrical resistivity shows a power law with an exponent of 7/2 at low temperatures. Investigation of the bulk magnetic properties reveals an extrapolated saturation magnetization of $\rm 5.0\,\mu_B/fu$ at 0~K. The films on $\rm Al_2O_3$ show an in-plane uniaxial anisotropy, while the epitaxial films are magnetically isotropic in the plane. Measurements of the X-ray magnetic circular dichroism of the films allowed us to determine element specific magnetic moments. Finally we have measured the spin polarization at the surface region by spin-resolved near-threshold photoemission and found it strongly reduced in contrast to the expected bulk value of 100\%. Possible reasons for the reduced magnetization are discussed.
\end{abstract}
\maketitle
\section{Introduction}
Highly spin polarized currents and hence materials with a high spin polarization at their surface are a crucial precondition for the fabrication of spintronic devices.\cite{PRI98} In this context the most interesting materials are half-metals. Such ferromagnets have only electrons of one spin direction at their Fermi edge. This property has been predicted for a number of materials. Photoemission\cite{PAR98} and tunneling\cite{BOW03} experiments carried out on $\rm La_{2/3}Sr_{1/3}MnO_3$ showed indeed half-metallicity at low temperatures. Unfortunately, the spin polarization at the Fermi level quickly vanishes at higher temperatures, which prohibits the use in applications.

Further candidates for half-metals can be found in the group of the half- and full-Heusler compounds. \cite{HEU03,deG83,GAL02a} Many of those materials have a Curie temperature considerably above room temperature. Half-metallicity has been experimentally verified in NiMnSb bulk samples.\cite{Han90} But surface sensitive techniques on NiMnSb thin films failed to demonstrate half-metallicity.  The successful deposition of epitaxial thin films has been reported for a number of Heusler alloys\cite{GEI02,KUB05,WAN05} and considerable progress in the preparation of tunneling magnetoresistance (TMR) elements with Heusler alloy electrodes has been made.\cite{KAM04,MAR05a,SAK06} While at low temperatures and low bias voltages large TMR effects have been achieved recently, the reported values at room temperature are not higher than the corresponding values obtained from conventional ferromagnets.\cite{SAK06}

It is not straightforward that the surface region of a Heusler alloy will show the same spin-polarization as the bulk material. In principle, extrinsic as well as intrinsic mechanisms may reduce the spin polarization at the surface. Among the extrinsic mechanisms, chemical and magnetic disorder have been discussed to play a relevant role.\cite{Rap02,Tok01} Intrinsic mechanisms, that take place also at perfectly ordered surfaces, are for example magnetic fluctuations and modifications in the surface band structure with respect to the bulk one. Band structure calculations for a number of Heusler alloys show a reduced spin polarization at the surface layers and a dependence on the terminating layer.\cite{GAL02} In a recent publication, Kolev \emph{et al.}\cite{Kol05} have investigated the (100) surface of NiMnSb using spin-resolved appearance potential spectroscopy and found a significantly reduced spin asymmetry with respect to calculations based on the bulk electronic structure. Interestingly, the spin polarization results are reduced over the whole energy range of the local density of states. Moreover, the authors could exclude chemical disorder, structural defects at the surface, as well as overall stoichiometric disorder as responsible mechanisms for the observed reduction of spin polarization. In another paper Wang \emph{et al.}\cite{WAN05} have studied single-crystalline Co$_2$MnSi films and found for the spin polarization of electrons at the Fermi level a maximal value of 12$\%$ measured with spin-resolved photoemission. They attribute this large discrepancy with the expected value of $100\%$ to partial chemical disorder in the Co$_2$MnSi lattice.

In this paper we investigate the Heusler alloy $\rm Co_2FeSi$. 
This material is especially interesting, since in a recent reinvestigation of its properties it was found to exhibit the highest Curie temperature of all materials and fulfilling the Slater-Pauling rule for half-metallic ferromagnets. \cite{GAL02a,WUR05} The value for the saturation magnetization in these samples is significantly higher than values reported in earlier publications.\cite{NIC79,BUS83} 
Interestingly, LDA band structure calculations fail to reproduce this value, instead an LDA+U formalism is required.\cite{GAL02a,WUR05} This emphasizes the importance of  electron correlations in Heusler alloys with a high number of valence electrons. Important in the context of spintronic applications is the fact that these LDA+U calculations predict a half-metallic band structure. 
The bulk samples showing the high values of the saturation magnetization had been annealed at 1300 K for 24 days. Such a procedure is not transferable to thin film 
deposition and therefore optimization of thin film deposition of this material is an important task, on which few results are published up to now. 
The deposition of very thin $\rm Co_2FeSi$ films on GaAs by molecular beam epitaxy has been reported recently.\cite{HAS05} These samples were grown at low temperatures in order to avoid interdiffusion with the reactive GaAs surface. First TMR elements have been fabricated, using sputtered, A2-oriented electrodes.\cite{TEZ06} The same paper mentions the deposition of $L2_1$ ordered films, but further analysis on these films has not been published.

Here we report the successful growth and full structural and magnetic characterization of $\rm Co_2FeSi$ thin films. As the electronic states at the interface of the tunneling barrier will determine the effective spin-polarization we use magnetic methods sensitive to different length scales due to their different information depths. The integral magnetic properties were determined with a  superconducting quantum interference device (SQUID). X-ray magnetic circular dichroism (XMCD) was used to detect element-specific magnetic moments. The spin polarization of the surface layers was investigated by spin-resolved near-threshold photoemission.

\section{Deposition and Structure}
We deposited $\rm Co_2FeSi$ thin films on $\rm Al_2O_3\,(11\bar{2}0)$ and MgO (100) substrates by RF magnetron sputtering. The stoichiometric targets were fabricated as reported in Ref.~\onlinecite{WUR05}. The base pressure of the deposition chamber was $5\times 10^{-8}\,\rm mbar$. We found that for both substrates best results with respect to crystallinity were obtained with an Ar pressure of $10^{-2}\,\rm mbar$ and a substrate temperature of 700~$\rm^\circ C$. The resulting deposition rate was about 5~\AA/s.{} After deposition the films were covered with 4~nm of Al at a temperature of 350~$\rm^\circ C$ in order to prevent oxidation.

\begin{figure}
\includegraphics[height=\columnwidth,angle=270]{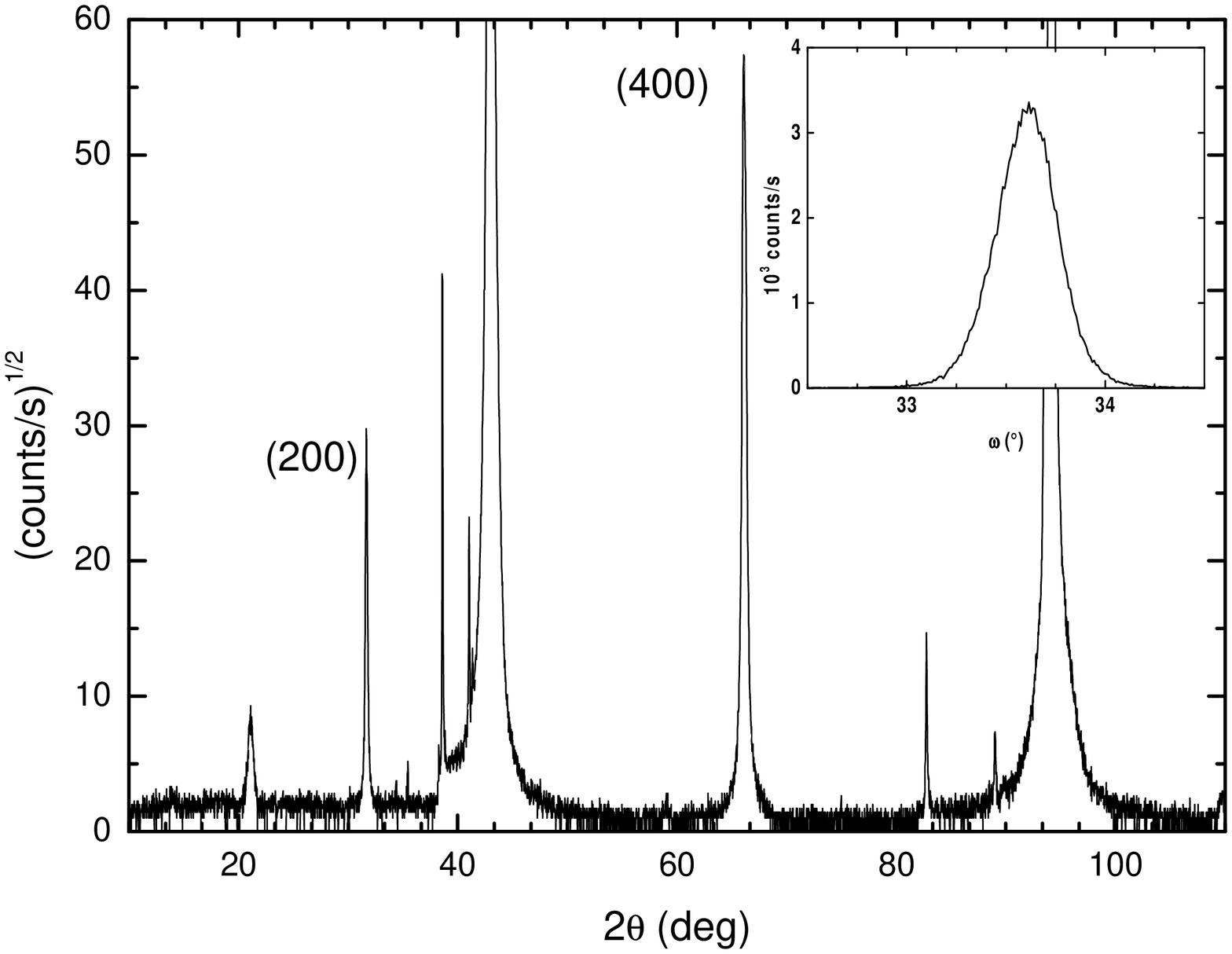}
\caption{$\theta$-$2\theta$-scan of a $\rm Co_2FeSi$ film deposited on MgO (100). The inset shows an $\omega$-scan of the (400)-reflection. Non indexed lines are due to the substrate and impurities of the anode.\label{braggmgo}}
\end{figure}

\begin{figure}
\includegraphics[height=\columnwidth,angle=270]{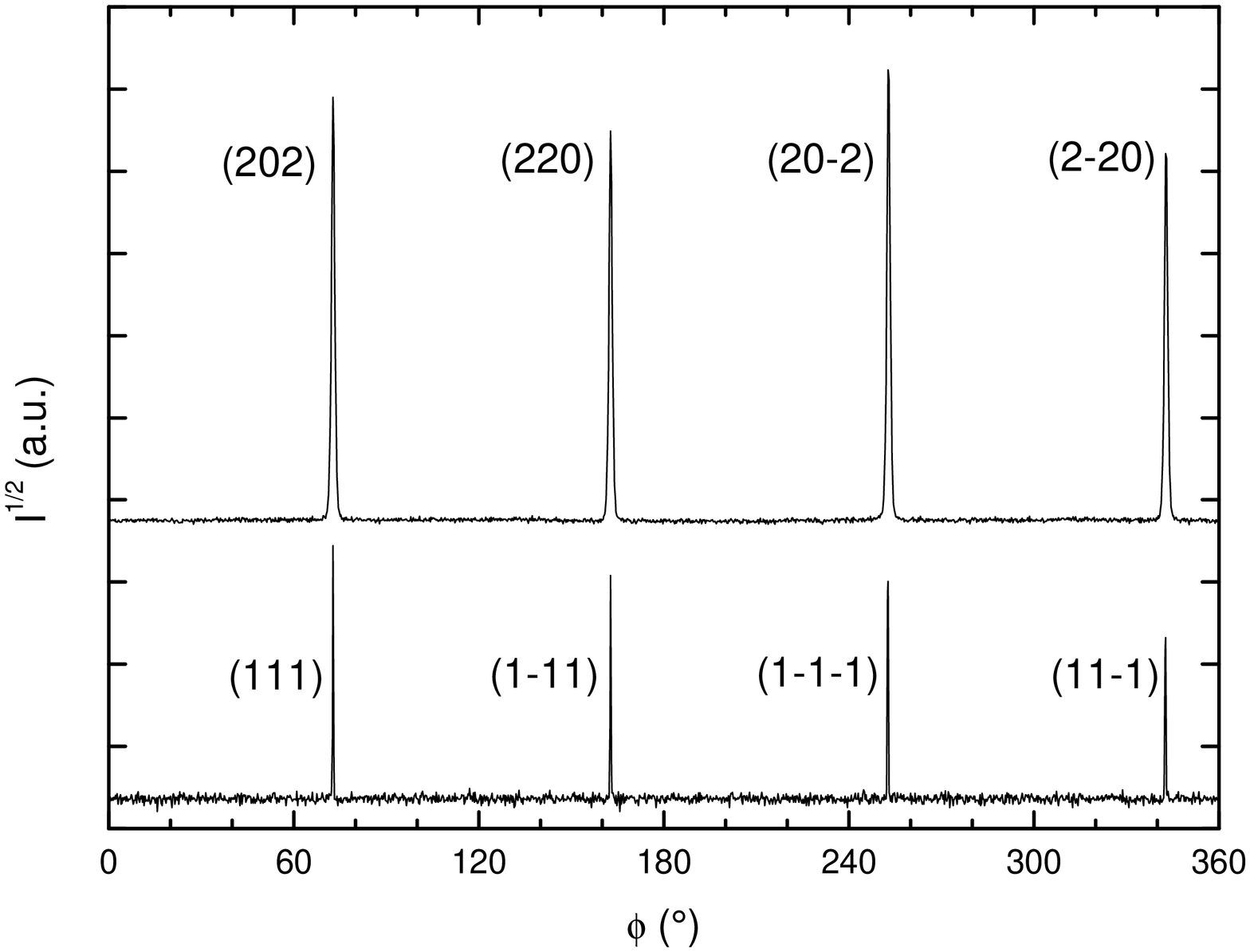}
\caption{$\phi$-scans of a (100)-oriented $\rm Co_2FeSi$ thin film (top) and the MgO (100) substrate.\label{mgophi}}
\end{figure}

Structural analysis was carried out with a Phillips X'Pert two-circle and a STOE four-circle diffractometer. Figure \ref{braggmgo} shows a Bragg scan of a $\rm Co_2FeSi$ film deposited on MgO. The (200) and (400) reflections of the film are clearly visible, while impurity phases cannot be detected. The corresponding rocking curve shown in the inset of Fig. \ref{braggmgo} has a width of $0.3^\circ$, which evidences a very good out-of-plane growth. The $\phi$-scan in Fig.~\ref{mgophi} reveals epitaxial growth. The (010) film plane is rotated by $45^\circ$ with respect to the (010) substrate plane.

\begin{figure}
\includegraphics[height=\columnwidth,angle=270]{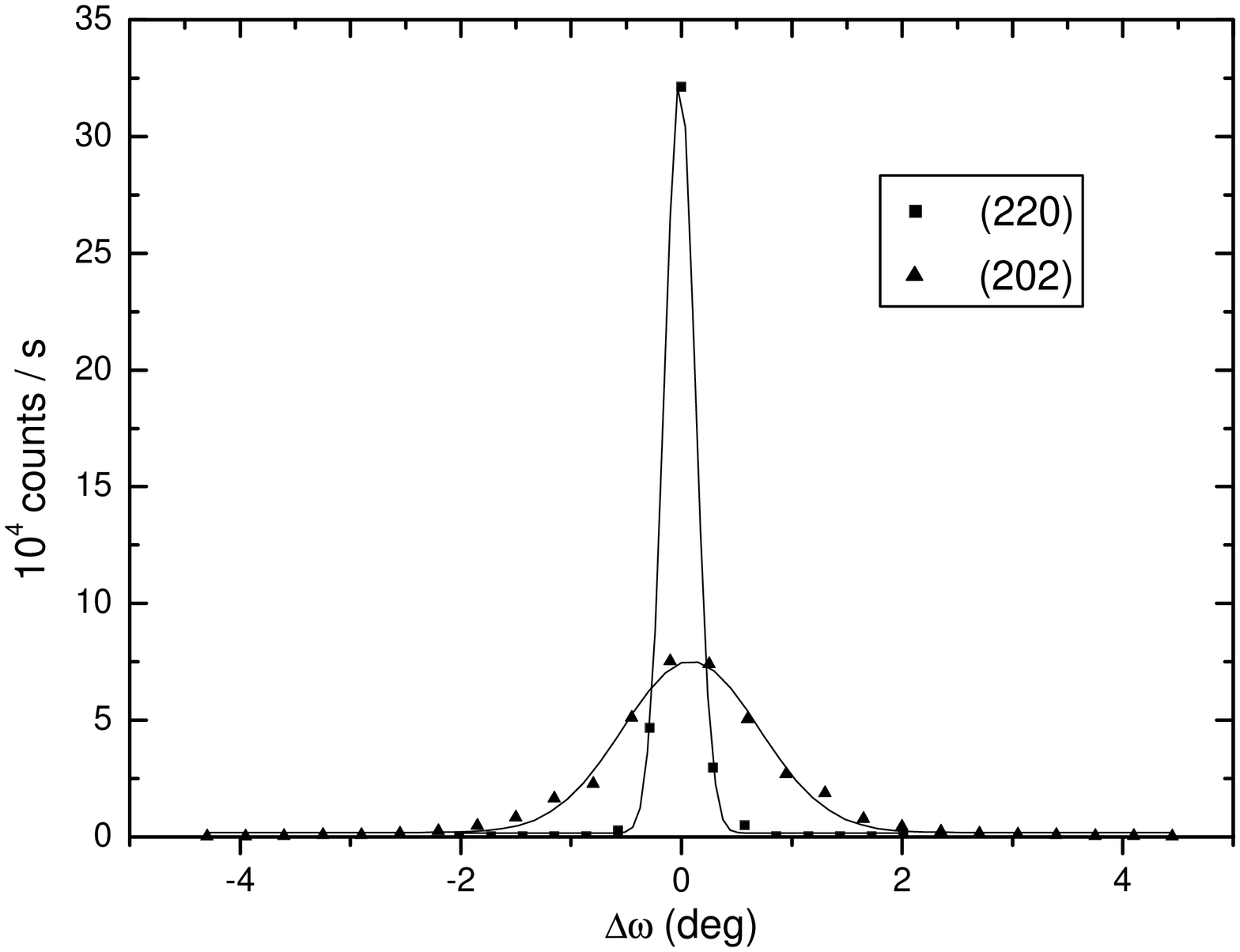}
\caption{$\omega$-scans of the specular (220) reflection and the off-specular (202) reflection of a film deposited on sapphire. The area below these two curves is comparable. \label{integrated_intensities}}
\end{figure}

A different growth orientation is observed on $\rm Al_2O_3$ substrates. Here the [110] direction is perpendicular to the surface. The out-of-plane rocking curves have a width of $0.1^\circ$. The Bragg reflections of thinner films show finite size oscillations, which indicates a flat film surface. For these films we find three epitaxial domains. The preferred orientation, which accounts for over 90\% of the scattered intensity, is $\rm (1\bar{1}0)_{film}\,||\,(0001)_{sub}$. For this alignment the misfit between film and substrate is very large, which may be the cause for the different competing in-plane domains.

The lattice constant for all films was determined to be 5.64~\AA{}, which was also found in bulk samples. Further scans of the reciprocal space show the presence of (111) and (311) reflections, which reveals that the films grow in the ordered $\rm L2_1$ Heusler structure. 
In principle the intensity of these reflections allows the 
determination of the degree of atomic disorder, as 
it is done in standard Rietveld refinements of powder patterns.
However, the thin film geometry leads to additional complications. 
Firstly there are typically much fewer reflections experimentally accessible than in powder samples due to the geometry of the four circle diffractometer. Secondly size and orientation effects are 
stronger. The film thickness may be very small against the lateral size of an island (or vice versa for columnar growth) leading to finite size 
related peak broadening of the respective reflections. 
Also the rocking curve width can depend strongly on the selected reflection. For comparison with calculations one can therefore not use the maximum count rates of the reflections but needs integrated
total intensities.
We compared our measured values with the intensities obtained from PowderCell simulations for different types and degrees of disorder. For films grown on MgO we need to introduce a disorder of 15-20\% between Co and Si sites to reproduce the experimental intensities. 
Despite the presence of several domains we find for films grown on $\rm Al_2O_3$ nearly the same integrated intensity for equivalent reflections (see Fig.~\ref{integrated_intensities}). Therefore we could estimate the disorder in these films as well and found the same amount of Co-Si site disorder. Due to the similar form factor of Fe and Co, disorder between these atoms is not accessible in standard x-ray analysis. 
The estimated values above will not necessarily be equivalent to a true atomic disorder in the thin film sample. While the formation of 
Co-Si antisites is expected to be highly unfavourable \cite{Pic04},  sputtered atoms are sufficiently energetic to introduce such defects.
Also ordered domains which are smaller than the coherence length will average out and result in a strongly reduced scattering intensity
of the ordering reflections.

\section{Macroscopic magnetic and transport properties}
\begin{figure}
\includegraphics[height=\columnwidth,angle=270]{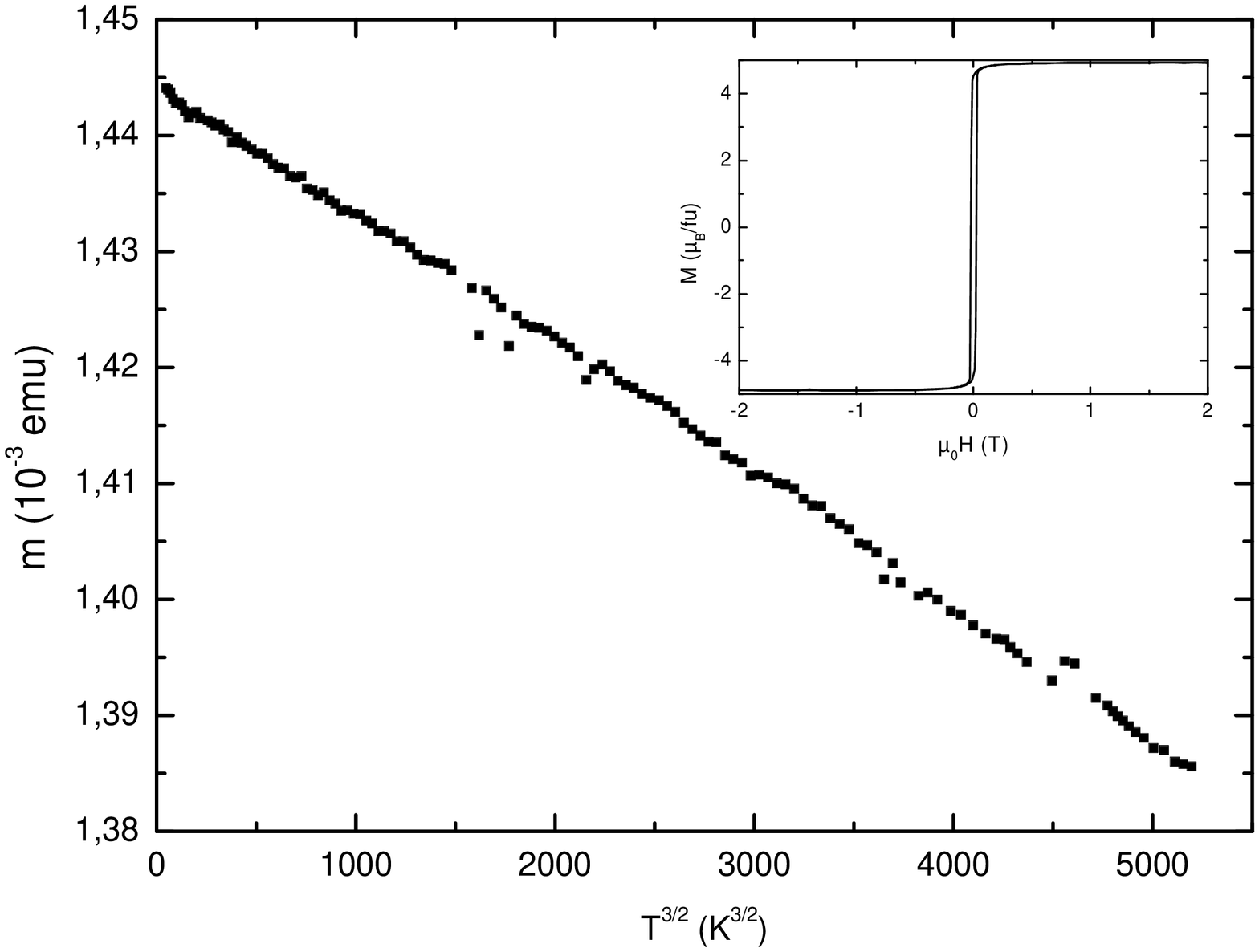}
\caption{Temperature dependence of the saturation magnetization of a $\rm Co_2FeSi/MgO$ film. The inset shows a hysteresis loop at 4~K.\label{hysteresis}}
\end{figure}

\begin{figure}
\includegraphics[height=\columnwidth,angle=270]{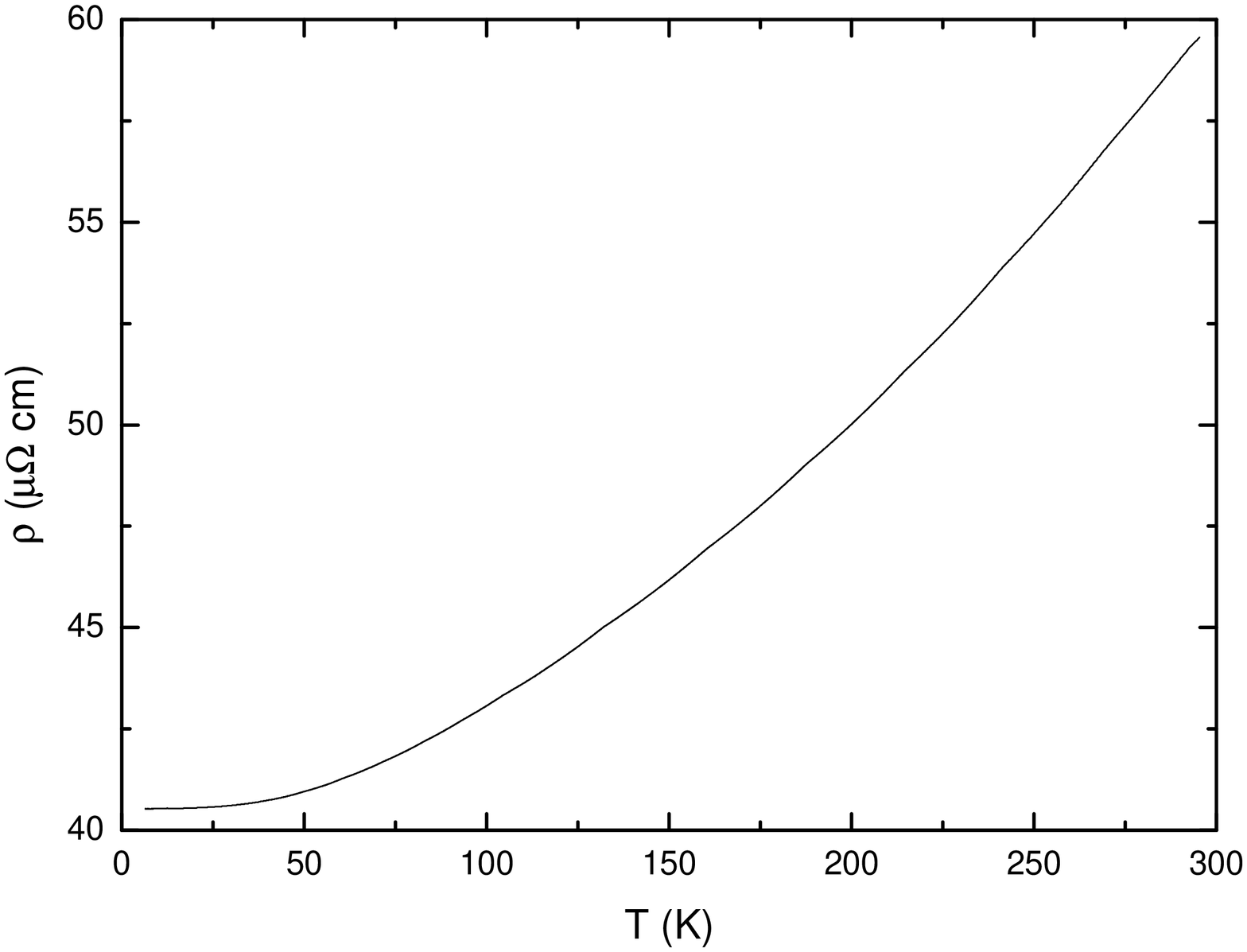}
\caption{Temperature dependence of the film resistivity. \label{RvsT}}
\end{figure}

\begin{figure}
\includegraphics[height=\columnwidth,angle=270]{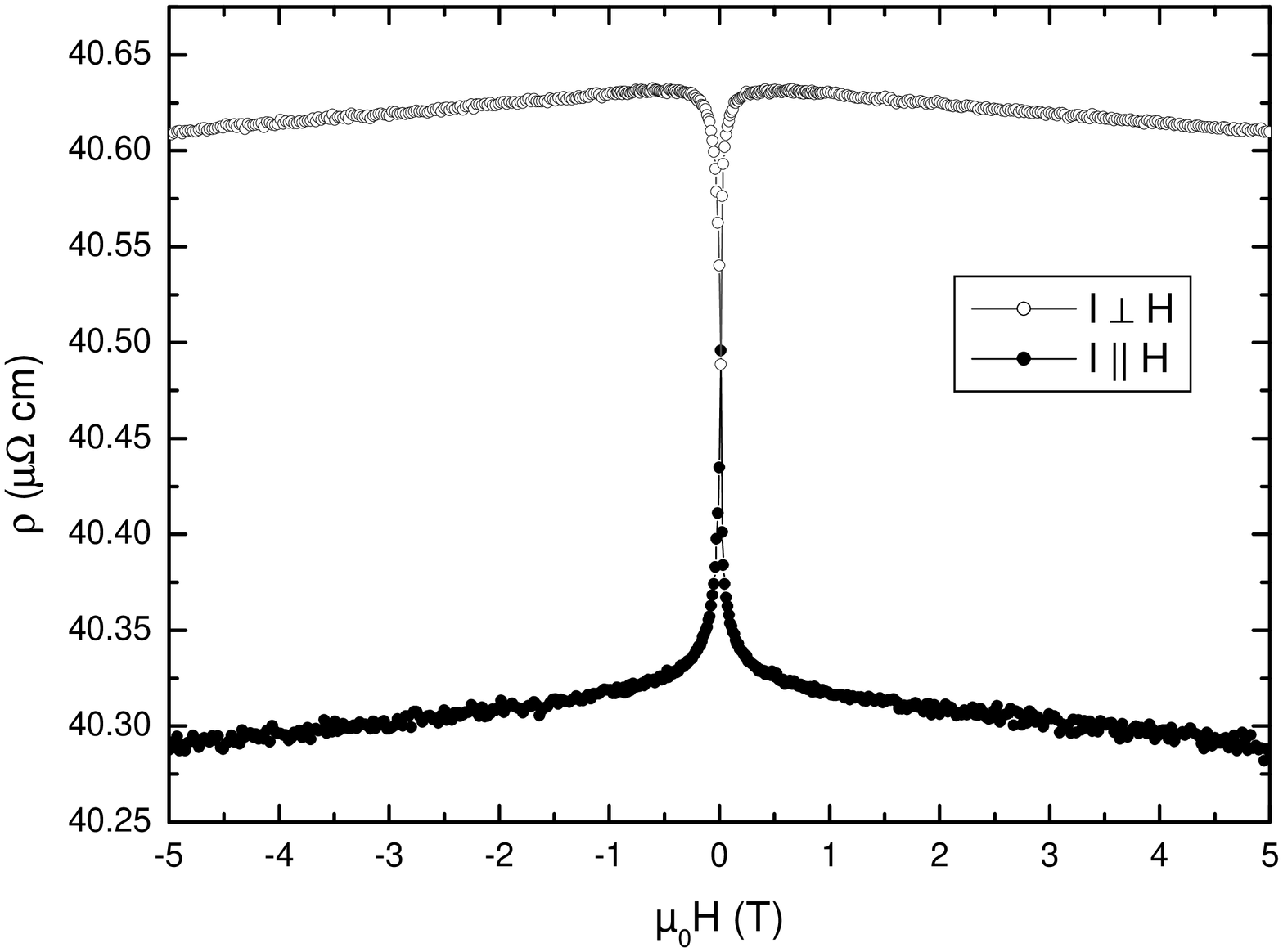}
\caption{Magnetoresistance of $\rm Co_2FeSi$ at a temperature of 4~K. \label{RvsB}}
\end{figure}

The bulk magnetization data was acquired with a Quantum Design SQUID. The results are presented in Fig.~\ref{hysteresis}. The saturation magnetization is independent on the chosen substrate and can be extrapolated to $\rm 5.0\,\mu_B/fu$ at 0~K. Magnetization data for different sample alignments with respect to the external field show an uniaxial anisotropy with the easy axis along the [110] direction. For the $\rm Co_2FeSi/Al_2O_3$ films we find $H_{c,[110]}=15\,\rm mT$ and $H_{c,[100]}=10\,\rm mT$. The films deposited on MgO show a much smaller anisotropy, here the values for the coercitive fields are 7.5~mT and 6.8~mT, respectively. The temperature dependence of the saturation magnetization shown in Fig.~\ref{hysteresis} exhibits a $T^{3/2}$ behavior over the whole investigated interval. This is in accordance to the classical spin wave picture for isotropic ferromagnets. 
In bulk polycrystalline material the temperature dependence is stronger ($\propto T^{1.8}$) and closer to the quadratic behavior of Fe bulk material. This difference between the bulk and film temperature dependence can result from the cutoff of spin waves due to the limited film thickness.\cite{KOB02}

Electrical transport measurements were carried out using the van der Pauw method. Figure \ref{RvsT} demonstrates that over the accessible temperature range our films show metallic behavior. The values for residual resistance ratio [$R(300\,{\rm K})/R(4\,{\rm K})=1.5$] and residual resistivity ($\rm\rho_0\approx 40\,\mu\Omega cm$) are comparable to values found in thin films of other Heusler alloys.\cite{MOO94,SIN04,GEI02} We can fit our data with a $T^{7/2}$ law below 70~K, at higher temperatures we find $(\rho-\rho_0) \propto T^{1.65}$. These exponents are to be considered as effective values resulting from several scattering mechanisms.
For conventional ferromagnets one would expect a $T^2$ caused by coherent one-magnon scattering processes. But processes like incoherent scattering\cite{RIC79} or s-d-scattering\cite{GOO63} may yield different exponents.  For half-metallic ferromagnets other scattering mechanisms have to be considered. In a rigid band model the absence of minority states yields a $T^{9/2}$ dependence.\cite{KUB72} For a non-rigid band a $T^3$ dependence was proposed.\cite{FUR00} However, the above models do not take into account the detailed shape of the Fermi surface, 
which can modify the temperature dependence.

The magnetoresistance data was measured with the film surface parallel to the external field and sample current parallel or perpendicular to the field. Figure \ref{RvsB} shows curves at a temperature of 4~K.  
We observe an anisotropic magnetoresistance effect with respect to the current direction, which implies the presence of spin-orbit coupling in our films. The spontaneous resistivity anisotropy 
$(\rho_\parallel - \rho_\perp)/(1/3\cdot\rho_\parallel +2/3\cdot \rho_\perp)\approx -0.8\%$ is small and negative. This effect is dominating the low field magnetoresistive response. After magnetic saturation the resistivity shows a linear dependence on the applied field. At 4~K the slope has a value of $\rm 4.5\,n\Omega cm/T$. It increases up to $\rm 25\,n\Omega cm/T$ at 300~K. These results are comparable to conventional ferromagnets.\cite{RAQ02} Therefore we must assume a significant contribution of spin-flip scattering to the magnetoresistance. 
  
\section{Microscopic magnetization (XMCD)}
\begin{figure}
\includegraphics[width=0.9\columnwidth]{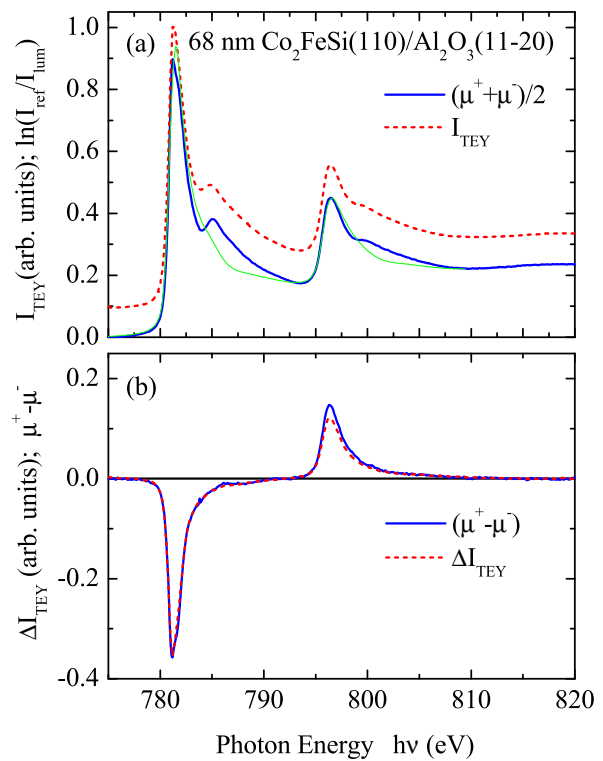}
\caption{(Color online)
Co XAS (a) and XMCD (b) spectra obtained by transmission measurements (full lines) and as determined by TEY (dashed lines) at 300~K of a 68~nm thick Co$_2$FeSi(110) Heusler alloy film grown on Al$_2$O$_3$(11$\bar{2}$0) and capped by 4~nm Al. The TEY XAS spectra were normalized to the value of the transmission XAS value at the $L_3$ maximum. The thin green line is an elemental Co reference.}
\label{Figure_xmcd_1_Co}
\end{figure}

\begin{figure}
\includegraphics[width=0.9\columnwidth]{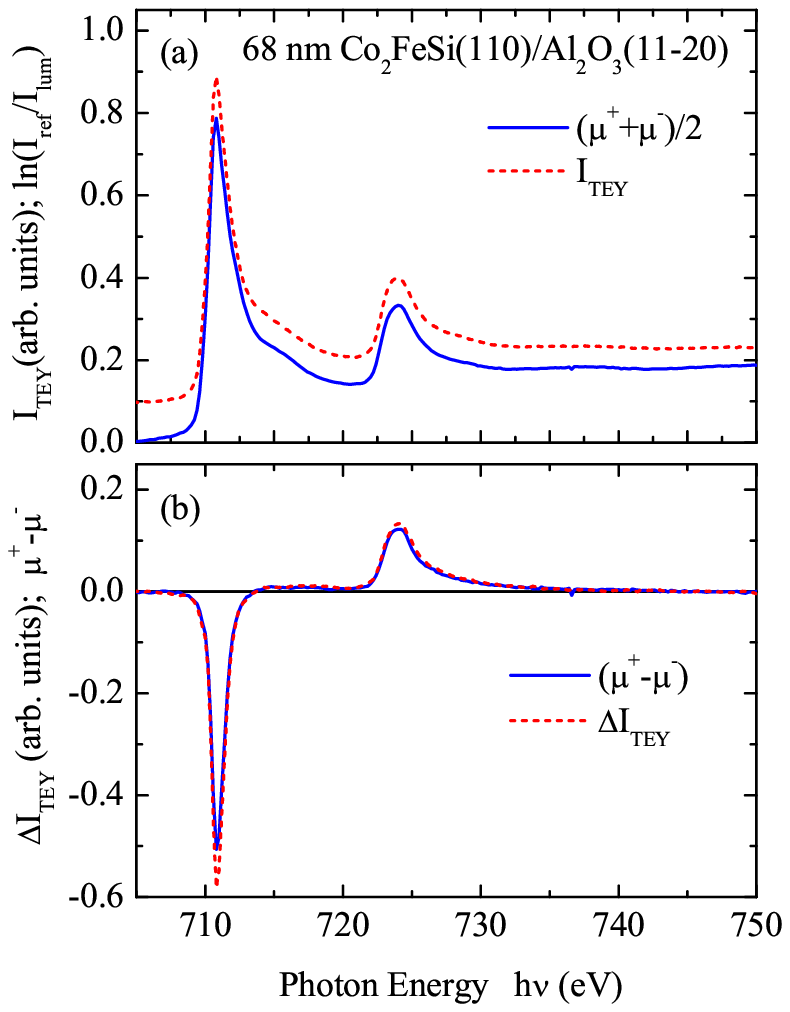}
\caption{(Color online)
Fe XAS (a) and XMCD (b) spectra obtained by transmission measurements
(full lines) and as determined by TEY (dashed lines) at 300~K
of
a 68~nm thick Co$_2$FeSi(110) Heusler alloy film grown on Al$_2$O$_3$(11$\bar{2}$0)
and capped by 4~nm Al. The TEY XAS spectra were normalized to the value of the
transmission XAS value at the $L_3$ maximum.
}
\label{Figure_xmcd_1_Fe}
\end{figure}

\begin{figure}
\includegraphics[width=0.9\columnwidth]{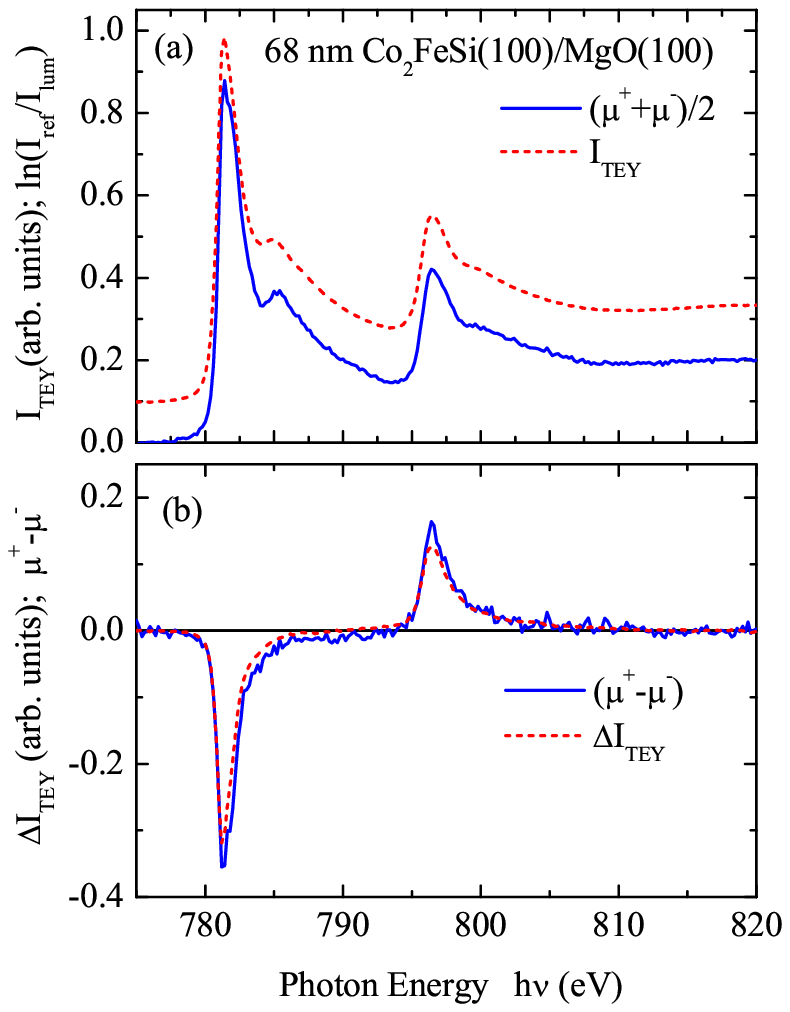}
\caption{(Color online)
Co XAS (a) and XMCD (b) spectra obtained by transmission measurements
(full lines) and as determined by TEY (dashed lines) at 300~K
of
a 68~nm thick Co$_2$FeSi(100) Heusler alloy film grown on MgO(100)
and capped by 4~nm Al. The TEY XAS spectra were normalized to the value of the
transmission XAS value at the $L_3$ maximum.
}
\label{Figure_xmcd_2_Co}
\end{figure}

\begin{figure}
\includegraphics[width=0.9\columnwidth]{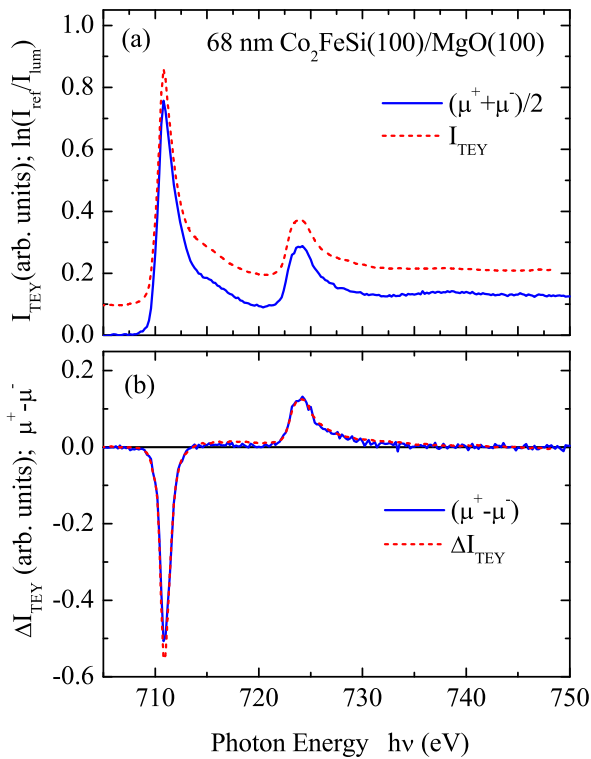}
\caption{(Color online)
Fe XAS (a) and XMCD (b) spectra obtained by transmission measurements
(full lines) and as determined by TEY (dashed lines) at 300~K
of
a 68~nm thick Co$_2$FeSi(100) Heusler alloy film grown on MgO(100)
and capped by 4~nm Al. The TEY XAS spectra were normalized to the value of the
transmission XAS value at the $L_3$ maximum.
}
\label{Figure_xmcd_2_Fe}
\end{figure}

\begin{table}
\caption{Element specific moments  for Fe and Co in the Co$_2$FeSi(110) films grown on Al$_2$O$_3$(11$\bar{2}$0) as derived from the sum rules from total electron yield (TEY) and transmission (TM) data. The specific number of $d$-holes were set to $N_d({\rm Co})=2.49$, $N_d({\rm Fe})=3.39$. Error bars for the XMCD values, excluding errors of $N_d$ or $P$, are $\pm 0.1\mu_{\rm B}$ for the spin moment and $\pm 0.02\mu_{\rm B}$ for the orbital moment. The summarized magnetic moment per formula unit $\mu_{\rm sum}=2\mu_{\rm Co}+\mu_{\rm Fe}$ is compared to the magnetic moment per formula unit measured by SQUID magnetometry at 300~K. The values from the XMCD experiment are approximately 10\% larger. The main uncertainty results from the assumed number of d-holes, which is probably not exactly the same as in elemental Co. Furthermore, the XMCD data does not take into account interstitial moments and the moment at Si sites, which have theoretically a magnitude of $0.2 \rm \mu_B$ and $0.1 \rm \mu_B$ respectively, and are aligned antiparallel to the Fe and Co moments.\cite{WUR05} Considering the general accuracy of the sum-rule analysis, XMCD and magnetometry data agree with each other.}
\label{Table1}
\begin{tabular}{lccc}
\hline
\hline
Al$_2$O$_3$(11$\bar{2}$0)&Co   &Fe   &Co$_2$FeSi\\
\hline
$\mu_{\rm spin}$(TEY)($\mu_{\rm B}$)    &1.13        &2.47    &4.73        \\
$\mu_{\rm spin}$(TM)($\mu_{\rm B}$)    &1.25        &2.43     &4.93        \\
\\
$\mu_{\rm orb}$(TEY)($\mu_{\rm B}$)    &0.14        &0.10     &0.38        \\
$\mu_{\rm orb}$(TM)($\mu_{\rm B}$)    &0.11        &0.07      &0.29        \\
\\
$\mu_{\rm sum}$(TEY)($\mu_{\rm B}$)    &1.27        &2.57      &5.11        \\
$\mu_{\rm sum}$(TM)($\mu_{\rm B}$)    &1.36        &2.50       &5.22        \\
\\
$\mu_{\rm SQUID}$($\mu_{\rm B}$)    &        &       &4.8        \\
\hline
\hline
\end{tabular}
\end{table}

\begin{table}
\caption{Element specific moments  for Fe and Co in
the Co$_2$FeSi(100) films grown on MgO(100)
as derived from the sum rules (see Tab.~\ref{Table1}).
}
\label{Table2}
\begin{tabular}{lccc}
\hline
\hline
MgO(100)&Co   &Fe   &Co$_2$FeSi\\
\hline
$\mu_{\rm spin}$(TEY)($\mu_{\rm B}$)    &1.07        &2.46    &4.60        \\
$\mu_{\rm spin}$(TM)($\mu_{\rm B}$)    &1.28        &2.46     &5.02        \\
\\
$\mu_{\rm orb}$(TEY)($\mu_{\rm B}$)    &0.04        &0.05     &0.13        \\
$\mu_{\rm orb}$(TM)($\mu_{\rm B}$)    &0.13        &0.12      &0.38        \\
\\
$\mu_{\rm sum}$(TEY)($\mu_{\rm B}$)    &1.11        &2.51      &4.73        \\
$\mu_{\rm sum}$(TM)($\mu_{\rm B}$)    &1.41        &2.58       &5.40        \\
\\
$\mu_{\rm SQUID}$($\mu_{\rm B}$)    &        &       &4.8        \\
\hline
\hline
\end{tabular}
\end{table}

Using X-ray magnetic circular dichroism (XMCD) in the X-ray absorption spectroscopy (XAS) we gain quantitative information on element-specific spin and orbital magnetic moments. The XAS experiments were performed at the UE56/1 - SGM beamline at the German synchrotron light source BESSY II (Berlin). While the incident photon flux was monitored by a Au net, we simultaneously measured the transmission XAS signal and the total electron yield (TEY) of the photoemitted electrons.\cite{KAL06} TEY was measured via the sample current. While the transmission signal integrates absorption properties along the film normal and thus represents the bulk properties of the film, the TEY signal stems from the upper interface region limited by the electron escape depth. Therefore, the TEY signal reflects properties of the upper interface. The sample was shielded by a conducting tube on a positive bias voltage (100~V) in order to collect all electrons. For the transmission XAS signal the photon flux transmitted through the thin Heusler films was detected via X-ray luminescence in the Al$_2$O$_3$ and MgO substrates.\cite{HUA04} The light intensity in the visible wavelength range (VIS) escaping at the substrate edge was measured by a GaAs - photodiode. The GaAs - photodiode was protected by a transparent cap layer in order to prevent detection of X-ray fluorescence light from the sample surface. An external magnetic field of 1.6~Tesla, which is sufficiently large to saturate the sample magnetization, was applied perpendicular to the film surface. We kept the X-ray polarization constant and flipped the magnetic field to determine the XMCD-signal. XMCD was measured at room temperature. Samples were capped by a protective 4~nm thick Al layer for these experiments.

Figures \ref{Figure_xmcd_1_Co} and \ref{Figure_xmcd_1_Fe} compare the transmission XAS and the X-ray absorption at the characteristic Co and Fe edges, respectively, as determined by TEY for a 68~nm thick Co$_2$FeSi(110) Heusler alloy film grown on Al$_2$O$_3$(11$\bar{2}$0). Assuming that the luminescence signal of the substrate $I^{\pm}_{lum}$ is proportional to the transmitted X-ray intensity, the X-ray absorption coefficient $\mu$ can be calculated using the equation $\mu^{\pm}(h\nu)=-ln[I^{\pm}(h\nu)/I_{\rm ref}(h\nu)]/d$, where $d$ is the thickness of the film. The reference spectra $I_{\rm ref}(h\nu)$ was measured by the bare substrate crystal and found to increase linearly with the photon energy. $I_{ref}(h\nu)$ was then normalized at the pre-edge region of the corresponding element (equivalent to an infinitely large penetration depth). The simultaneously measured TEY absorption spectra are shown in Figs.~\ref{Figure_xmcd_1_Co} and \ref{Figure_xmcd_1_Fe} for comparison. After subtracting the background signal the TEY-XAS spectra were multiplied by a constant factor in order to achieve $(I_{\rm TEY}^++I_{\rm TEY}^-) = \mu^++\mu^-$ at the $L_3$ maximum. The corresponding normalization factor was also applied to the XMCD signal in order to compare both signals. The TEY spectra show no oxidation features indicating that the Al layer protects the surface.

Transmission and TEY Co XAS signals (Fig.~\ref{Figure_xmcd_1_Co}a) show the typical features observed already for bulk Co$_2$CrAl and Co$_2$FeSi Heusler alloy.\cite{ELM03,WUR05} The general appearance of both spectra are similar. Even the step jump between pre- and post-edge intensity is almost equal, in contrast to the previously investigated case of Co$_2$Cr$_{0.6}$Fe$_{0.4}$Al/Al$_2$O$_3$ films.\cite{KAL06} This indicates similar values for the number of $d$-holes in the bulk and at the interface. Compared to the XAS signal of pure Co, the signal drops less steep after the $L_{2/3}$ edge and shows an extra maximum at about 4~eV above the edges.\cite{ELM03,WUR06} This extra maximum is much more pronounced for the $L_3$ edge than for the $L_2$ edge because of the typical life-time broadening. The extra peak was explained by a hybridization of Co d-band states with sp-states of the main group element. We have observed that this extra peak vanishes for disordered or selectively oxidized Heusler alloys.\cite{ELM03} The fact that the extra peak is more pronounced for the transmission signal compared to TEY might indicate that the bulk of the film shows a higher degree of local atomic order than the surface region.

The XMCD signals for transmission and TEY do not show any pronounced differences for films grown on Al$_2$O$_3$. Element specific magnetic moments were derived by a sum-rule analysis assuming numbers of $d$-holes $n_d(\rm Fe)=3.4$, $n_d(\rm Co)=2.5$ as reported for the pure elements.\cite{NAK99} Theoretically predicted values of $n_d$ in the Heusler compound deviate slightly from the pure element data for the case of an LDA based calculation: $n_d(\rm Fe)=3.5$, $n_d(\rm Co)=2.3$.\cite{Galanakisprivate} Smaller values in particular for Co were reported for a calculation considering electron correlation (LDA+U) ($n_d(\rm Fe)=3.2$, $n_d(\rm Co)=1.8$).\cite{WUR05} An experimental determination of $n_d$ results from an integration of the spin-averaged absorption signal after subtraction of a step function.\cite{NAK99} A comparison with the corresponding value that we measured for a pure element reference sample (see Fig.~\ref{Figure_xmcd_1_Co}a) reveals that $n_d(\rm{Co})$ is almost equal for pure Co and Co in Co$_2$FeSi. We estimate that the systematic error due to background subtraction is less than 10~\% for this measurement. This result justifies the usage of Co and Fe bulk values for the determination of magnetic moments. 

We corrected the XAS spectra for saturation effects using the X-ray penetration depth as determined from the transmission signal and assuming an electron escape depth of $\lambda_e=25$~\AA{}.\cite{NAK99} Neglecting the saturation effect results in apparently smaller spin and orbital moments, much more pronounced for orbital moments. However, in the present case the correction is small because the elements are diluted in the Heusler alloy.\cite{KAL06} For calculation of the spin moment we neglected the magnetic dipolar asymmetry. This is justified since we did not find an angular dependence of the spin moment. The element specific moments are summarized in Tab.~\ref{Table1}. Note, that the error bars given for the spin moment consider only errors due to the integration of the spectra and not errors in the number of d-holes or in the polarization value of the X-ray light. Interestingly, for films grown on Al$_2$O$_3$ the surface and bulk magnetic moments do not show a difference within our error limits.

Similar results obtained for a 68~nm thick Co$_2$FeSi(100) Heusler alloy film grown on MgO(100) are shown in Fig.~\ref{Figure_xmcd_2_Co} and Fig.~\ref{Figure_xmcd_2_Fe}. Also for the (100) oriented films the extra peak in the Co absorption signal appears somewhat smaller in the TEY signal compared to the transmission signal. The difference is slightly more pronounced compared to the (110) oriented film. We also observe a significant difference in the Co XMCD signal.  While the asymmetry of the transmission signal is equal to the value measured for the (110) oriented film the TEY XMCD signal is significantly smaller both at the $L_3$ and $L_2$ edge. The sum rule analysis consequently results in smaller values for the surface magnetic moment compared to the bulk value (see Tab.~\ref{Table2}). Tentatively we attribute this effect to a less well ordered interface region at the Al cap layer of (100) oriented films compared to (110) oriented films. An even larger decrease of surface moment with respect to the bulk value was observed in the case of  Co$_2$Cr$_{0.6}$Fe$_{0.4}$Al/Al$_2$O$_3$(11$\bar{2}$0) films.\cite{KAL06}

The summarized magnetic moments $\rm\mu_{sum}$ per formula unit resulting from XMCD appear smaller than the integral value determined by magnetometry. A slightly larger value could be expected, since the antiparallel aligned interstitial and Si moments are not considered in the XMCD data. Moreover, the number of d-holes assumed here might be overestimated because theoretical values for $n_d$ are slightly smaller.\cite{WUR05,Galanakisprivate}

The absolute values of the element-specific magnetic moments are smaller than expected for a half-metallic material and smaller than those determined for the corresponding bulk material. A band-structure calculation considering electron correlation\cite{WUR05} predicts values of $\mu({\rm Fe})=3.3$~$\mu_{\rm B}$, $\mu({\rm Co})=1.5$~$\mu_{\rm B}$ and a magnetization of $\mu=6.0$~$\mu_{\rm B}$/fu. One expects an integer value for the number of $\mu_{\rm B}$/fu for a half-metallic Heusler alloy and indeed a magnetization of $\mu=6.0$~$\mu_{\rm B}$/fu was experimentally confirmed for bulk samples of Co$_2$FeSi.\cite{WUR05}

\section{surface spin polarization}

\begin{figure}
\includegraphics[width=4.15cm,keepaspectratio]{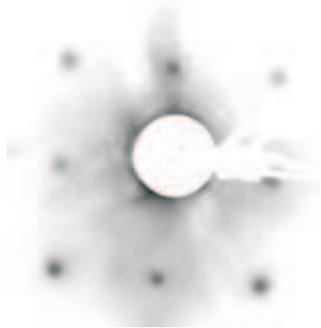}
\caption{LEED pattern of $\rm Co_2FeSi$/$\rm MgO$, taken at 318\,eV primary electron energy. The pattern demonstrates a clear fourfold symmetry. \label{LEED}}
\end{figure}

We now discuss the spin-resolved pho\-to\-emis\-sion spectra (SRPES) measured from our $\rm Co_2FeSi$/$\rm Al_2O_3$ and $\rm Co_2FeSi$/$\rm MgO$ thin films. The samples were mounted in an ultrahigh vacuum (UHV) chamber with a base pressure below $10^{-10}$\,mbar. Prior to the measurements, the Al protecting cap layer was removed under UHV conditions \textit{in situ} by sputtering with 500\,eV Ar$^{+}$ ions. Subsequently, the surface was prepared by repeated cycles of sputtering and prolonged annealing at 570\,K. Cleanliness was checked by means of Auger electron spectroscopy, geometrical order by means of low energy electron diffraction (LEED). Only the $\rm Co_2FeSi$/$\rm MgO$ films, being fully epitaxial, revealed a LEED pattern. A representative LEED pattern for such films is shown in Fig. \ref{LEED}. The picture was recorded at 318\,eV primary electron energy and demonstrates a clear fourfold symmetry in correspondence with the cubic bulk lattice.

\begin{figure}
\includegraphics[width=\columnwidth,keepaspectratio=true]{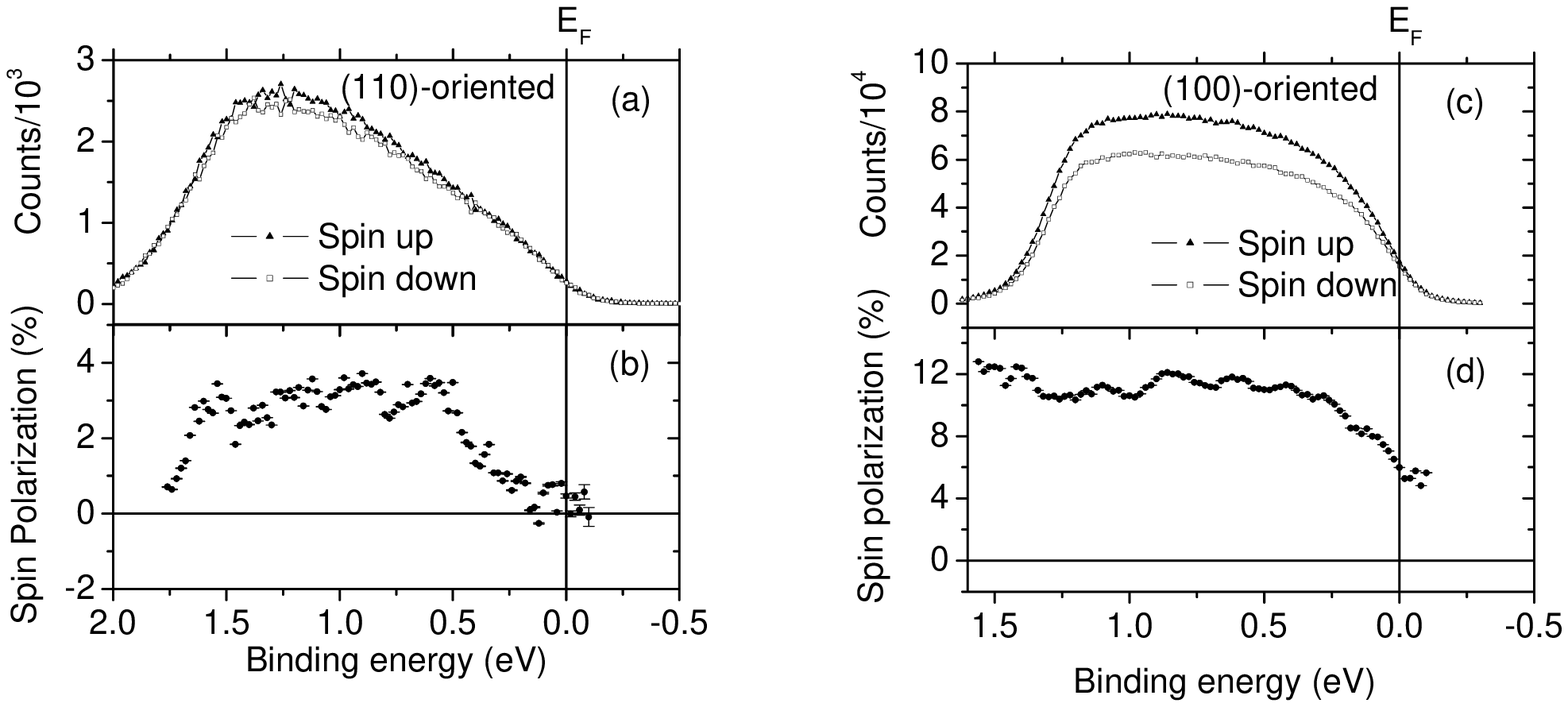}
\caption{ (a) and (c) Spin-resolved photoemission spectra of the
$\rm Co_2FeSi$/$\rm Al_2O_3$ and  $\rm Co_2FeSi$/$\rm MgO$ films.
The excitation source was the fourth harmonic of a Ti:Sapphire
oscillator, with photon energy of $6.0$\,eV (see text). Triangles
(squares) indicate majority (minority) spin spectra. (b) and (d)
Spin polarization calculated from the curves in (a) and (c),
respectively. \label{SRPES}}
\end{figure}

The SRPES were recorded at 300K using the s-polarized 4th harmonic (photon energy 5.9~eV) of a narrow band, pulsed Ti:Sapphire oscillator (Spectra-physics, Tsunami) created by sequential frequency doubling in two $0.2$\,mm thick beta barium borate (BBO) crystals. The incident light was focused onto the sample with an angle of 45$^{\circ}$ with respect to the surface normal.  The photoemitted electrons were analyzed in normal emission geometry by a commercial cylindrical sector analyzer (Focus CSA 300) equipped with an additional spin detector based on spin-polarized low-energy electron diffraction (Focus SPLEED). The achieved energy resolution is 150\,meV, the acceptance angle of the analyzer is $\pm 13^{\circ}$.  For the SRPES measurements the films were remanently magnetized by an external in-plane magnetic field. 
Figure \ref{SRPES} (a) and (c) show the SRPES recorded from the $\rm Co_2FeSi$/$\rm Al_2O_3$ and  $\rm Co_2FeSi$/$\rm MgO$ films, respectively. The spectra show the Fermi level ($E_F$) and a structureless behavior up to the low energy cutoff. This is in qualitative agreement with band structure calculations performed with LDA$+U$ and presented in Ref. \onlinecite{WUR05}. However, those calculations predict for binding energies between approximately 2\,eV and $0.5$\,eV a higher density of states for minority than for majority electrons, and a band gap for minority electrons in the energy range between $0.5$\,eV and $E_{F}$. One would expect a 100\% polarization between $0.5$\,eV and $E_{F}$ and a sudden drop to negative spin polarization for binding energies between 2\,eV and $0.5$\,eV.

However, in the measured spin up and spin down spectra shown in Figs. \ref{SRPES} (a) and (c) the majority-spin spectrum  has a higher intensity than the minority-spin spectrum over the whole energy range and for both films. This results in a positive spin-polarization, depicted in Fig. \ref{SRPES} (b) and (d). In more detail, the curves show a maximal spin polarization of $4\%$ for the $\rm Co_2FeSi$/$\rm Al_2O_3$ film and of $12\%$ for the $\rm Co_2FeSi$/$\rm MgO$ film, respectively. For energies close to $E_F$ we have $P(E_F)\approx0\%$ and $P(E_F)\approx 6\%$, respectively. These values are  drastically reduced with respect to the  expected $100\%$ in case of  half-metallic behavior as predicted in Ref. \onlinecite{WUR05}. Similar values have been reported previously by Wang \emph{et al.} for single-crystalline $\rm Co_2MnSi$ films.\cite{WAN05} In accordance with the results of Kolev \emph{et al.}\cite{Kol05} we also observe that the the discrepancy with theory is not restricted to $E_F$, but extends up to 1.5\,eV below $E_F$. Further studies of a Co$_2$FeSi/MgO film performed with spin- and time-resolved photoemission, have shown that the observed reduction of spin polarization is not confined to the outermost surface layer, but extends at least 4 -- 6~nm into the sample \cite{WUS06}.

In contrast the XMCD and SQUID experiments showed only a comparatively small reduction of the expected values. This discrepancy results from the particular band structure of this Heusler compound. The half metallic behavior of the perfectly ordered compound originates from a complete absence of spin minority states at $E_F$ and a small but nonvanishing density of majority states. As discussed for the similar Co$_2$MnSi compound a small degree of disorder can introduce spin minority states at $E_F$.
Especially Co antisite defects act in this direction and will have a dramatic effect on spin polarization at $E_F$.\cite{Pic04} We found in our XMCD experiments a pronounced double peak structure at the Co-edges and also clearly observed the $L2_1$ order peaks in the x-ray data.  Both observations are usually taken as proof for an $L2_1$-ordered sample. But from our analysis of the x-ray peak intensities we concluded on atomic site disorder. Here the surface has been treated by sputter cleaning, which can introduce additional 
atomic disorder. Obviously the disorder level seems to be sufficient to destroy half metallicity at $E_F$ in the photoemission experiment. However, XMCD and SQUID experiments probe integrated densities of states above and below $E_F$, respectively. The small density of states close to $E_F$ does not contribute much to the integral values and we find correspondingly only small reductions of the magnetic moments.

\section{Summary}
We have successfully deposited thin films of the potentially half-metallic material $\rm Co_2FeSi$, which grows epitaxially in the $\rm L2_1$ full Heusler structure. Despite the good crystal quality, the bulk value of the saturation magnetization is reduced in comparison with the value from polycrystalline bulk samples. This might be caused by an incomplete atomic order with respect to the $\rm L2_1$ structure in our films. Using methods sensitive to different length scales we find that the reduction is not homogeneous over the film thickness. Magnetic moments are lowest at the interface to the Al-cap layer as can be deduced from comparison of transmission and total electron yield signals in XMCD experiments.  
Using spin resolved photoemssion experiments we failed to establish a high spin polarization at $E_F$. Due to the high energy resolution these experiments 
will be more sensitive to disorder effects which introduce minority states at $E_F$. 
The unavoidable sputter cleaning of the sample, however, could introduce a higher degree
of disorder at the free surface. 

From our results we conclude that optimization of 
the atomic ordering will be the most important factor in order to establish a high spin polarization in $\rm Co_2FeSi$ films.

\begin{acknowledgments}
We would like to thank S. Cramm for support at BESSY. Furthermore we thank the DFG for their financial support through Forschergruppe 559.
\end{acknowledgments}

\end{document}